# Enhanced spin-orbit coupling in dilute fluorinated graphene

*Ahmet Avsar[1,2,*], Jong Hak Lee[1,2,*], Gavin Kok Wai Koon[1,2], and Barbaros Özyilmaz[1,2,3,†]*

[1]Centre for Advanced 2D Materials, National University of Singapore, 117542, Singapore,
[2]Department of Physics, National University of Singapore, 117542, Singapore,
[3]NanoCore, National University of Singapore, 117576, Singapore.

The preservation and manipulation of a spin state mainly depends on the strength of the spin-orbit interaction. For pristine graphene, the intrinsic spin-orbit coupling (SOC) is only in the order of few μeV, which makes it almost impossible to be used as an active element in future electric field controlled spintronics devices. This stimulates the development of a systematic method for extrinsically enhancing the SOC of graphene. In this letter, we study the strength of SOC in weakly fluorinated graphene devices. We observe high non-local signals even without applying any external magnetic field. The magnitude of the signal increases with increasing fluorine adatom coverage. From the length dependence of the non-local transport measurements, we obtain SOC values of ~ 5.1 meV and ~ 9.1 meV for the devices with ~ 0.005% and ~ 0.06% fluorination, respectively. Such a large enhancement, together with the high charge mobility of fluorinated samples (μ~4300 $cm^2$/Vs - 2700 $cm^2$/Vs), enables the detection of the spin Hall effect even at room temperature.

[†]Address correspondence to: barbaros@nus.edu.sg
[*]These authors contributed equally to this work.

For pristine graphene, the hopping of $\pi$ electrons between the two next nearest neighbor carbon atoms is the only source of the SOC. This is a second order process and gives rise to SOC in the order of only a few µeV [1–3]. Such a weak coupling limits the prospect of potential graphene-based spin field effect transistors [4]. Different approaches have been suggested to enhance the SOC of graphene. For example, the creation of a curvature in flat graphene is expected to significantly enhance the intrinsic SOC [1,3,5] as in the carbon nanotube case [6]. However, there is no well-developed method for the fabrication of such devices. Recently, the proximity effect at the interface between graphene and $WS_2$ layers has been shown to result in a SOC enhancement [7]. Also the hydrogenation of graphene has been shown to significantly enhance the SOC of graphene [8,9]. Unfortunately, chemisorbed hydrogen atoms can be easily detached at moderate temperatures and this makes devices less stable at ambient conditions [10,11]. The stability of devices can be improved with choose of other types of adatoms [12]. On this subject, fluorinated graphene is known to be thermodynamically more stable. Here, the carbon forms strong covalent bonds with fluorine, which is known as the most electronegative element [13]. This chemical bonding depends strongly on graphene doping and hence allows a new route to tailor the electronic and spintronics properties with a local gate similar to the one can be achieved with magnetic adatoms [14–16]. Furthermore, and unlike hydrogen, fluorine acts only as a weak resonant scatterer [17]. Therefore, higher charge mobility is expected in such weakly functionalized graphene. And lastly but most importantly, contrary to the hydrogen atoms, fluorine's own SOC is not negligible and is expected to cause a large SOC in even weakly fluorinated graphene system [2,17]. Irmer et al., has predicted a SOC strength larger than 10 meV in weakly fluorinated graphene which is ten-fold higher what is expected in hydrogenated graphene and comparable to the atomic spin-orbit interaction in carbon itself [17].

In this letter, we study the strength of spin-orbit interaction in dilute fluorinated graphene by measuring the non-local signals of Hall bar devices. Prior to transport measurements, Raman spectroscopy is utilized to determine the fluorine adatom coverage of graphene. We observe a large non-local signals which increases with increasing the coverage. By fitting the length dependence of the non-local signal, we estimate a SOC of up to 9.1 meV. The observation of such larger SOC strength in fluorinated graphene is in a good agreement with recent studies where a large SOC-induced band splitting has been predicted [2,17].

The device fabrication starts with employing the well-known micromechanical exfoliation method to obtain single layer fluorinated graphene on Si/SiO$_2$ wafers. However instead of using pristine HOPG graphite, we use ClF$_3$ treated graphite. By controlling the temperature and duration of the treatment, we can achieve the transition from insulating to conducting graphene with a non-destructive recovery (See Supplementary Information). This way we can prepare weakly fluorinated graphene flakes with different fluorine concentrations. The details of the synthesis is discussed in detail elsewhere [18]. Subsequently devices are fabricated using electron beam lithography technique followed by Au/Cr electrode deposition. Following the lift off process, a second electron beam step is performed to etch the graphene into Hall bar structures. Figure 1-a shows a typical optical image of a completed device. The longitudinal spacing between the electrodes ($l$) is varied from 2 μm to 4 μm while keeping the width of graphene channel fixed to 1 μm. Prior to the transport measurements, the homogeneity of the fluorine coverage over the graphene surface is checked with Raman mapping (Figure 1-b). All transport measurements are performed with a four terminal ac lock-in technique under room temperature and vacuum environment (pressure ~ 1 x 10$^{-7}$ mbar). Experiments are performed in two different measurement configurations (Figure 1c). In the conventional local four-terminal measurement configuration (Hall bar), a current of 1 µA flows between electrode 1 and electrode 6 and a local voltage drop is measured between electrode 2 (3) and electrode 4 (5). In the non-local measurement configuration (H bar), the current $I$ flows between the pair of electrode 2 and electrode 3, and a non-local voltage $V$ is recorded across the neighboring pair of electrode 4 and electrode 5. In total, we have characterized 5 dilute fluorinated graphene devices. Here we discuss two representative fluorinated devices at the maximum (device *FG2*) and minimum (device *FG1*) fluorination limits and one pristine graphene device (PG) as a control.

Figure 2-a shows the Raman spectra of the flakes. Having larger 2D intensity ($I_{2D}$) peak compared to G intensity ($I_G$) peak shows that all flakes are single layer [19]. The intensity of the defect-associated graphene D band ($I_D$) is absent in pristine graphene. The observation of higher $I_D$ and an emerging D´ peak in *FG2* compared to *FG1* indicates a higher fluorine coverage in the former [18]. The Raman intensity of the $I_D$ which is normalized to the $I_G$ allows us to determine the spacing between the fluorine atoms ($L_D$) and hence, the fluorine concentration ($n_{imp}$) by using the relation [20]

$$L_D^2 (nm^2) = (1.8 \pm 0.5) x 10^{-9} \lambda_L^4 (I_G/I_D)$$

and $$n_{imp}(cm^{-2}) = 10^{14}/(\pi L_D^2),$$

where $\lambda_L$ is the wavelength of the Raman laser which is 532 nm. The $I_D/I_G$ ratios for *FG1* and *FG2* give $L_D$ ~28 nm and 8 nm and $n_{imp}$ = 4 x $10^{11}$/cm² and 4.6 x $10^{12}$/cm² respectively. From the extracted $L_D$ values, fractions of the fluorination are estimated to be 0.005% and 0.06% for *FG1* and *FG2*, respectively from $\frac{3\sqrt{3}}{\pi}\left(\frac{a}{L_d}\right)^2 x100$. These imply that our devices are very weakly fluorinated and *FG2* has one order of magnitude higher fluorine coverage compared to *FG1*.

These devices are first characterized by using local charge transport experiments to confirm their homogeneity. Local charge transports are performed for top and bottom electrodes in a single Hall bar junction. Only junctions which show similar top and bottom local contributions are further used for non-local spin Hall effect (SHE) measurements. Figure 2-b shows the carrier concentration ($n$) dependence of the local resistivity ($\rho$) for *FG1*, *FG2* and pristine graphene devices. The resistivity of *FG2* is highest, followed by *FG1* and pristine devices. This is in a good agreement with the impurity concentration of the devices. The full width at half maximum is largest in *FG2* indicating that the sample has the lowest charge mobility. The field effect mobility of 2700 cm²/Vs, 4300 cm²/Vs and 7350 cm²/Vs are extracted for *FG2*, *FG1* and pristine graphene devices, respectively, by using $\mu = \frac{1}{e}\frac{d\sigma}{dn}$ where $\sigma = \frac{1}{\rho}$ and $n = (7.2 \times 10^{10}\ cm^{-2}\ V^{-1})(V_G - V_G^0)$. We note that the mobility of *FG2* device is even higher than that of the hydrogenated graphene device despite the larger adatom coverage in the former [8]. In fact, it is comparable to the pristine graphene-based spin valve devices with tunnel barrier fabricated on Si/SiO₂ substrate [21–23].

We now turn our attention to non-local transport measurements. We first characterize PG which does not have any functionalization treatment. The obtained non-local signal ($R_{NL}$) has comparable magnitude with the Ohmic leakage contribution, $R_{Ohmic} \sim \frac{4}{\pi}\rho\exp(-\pi L/w)$, and thus there is no indication of the SHE (Inset Figure 3-a and see supplementary information) [8,24,25]. However, we observe a $R_{NL}$ signal approximately ~ 12 times higher than the estimated $R_{Ohmic}$ in the *FG2* device for $l$ = 2.5 µm junction. Figure 3-b shows the $n$ dependence of $R_{NL}$ at different length values for device *FG2*. While the non-local signal decreases as the device length is increased, all the measured signals are almost an order of magnitude larger than the expected Ohmic contributions. Thus the presence of this non-local signal at room temperature and zero magnetic field suggests the enhancement of SOC [7,8,26]. The large enhancement in SOC gives rise to the generation and detection of spin currents via the SHE and inverse SHE respectively. In order to determine the important spin parameters in our devices, we study the length dependence of the non-local signal. Figure 3-c shows the $R_{NL}$ normalized by the local $\rho$ at n=1x10$^{12}$cm$^{-2}$. At zero applied field, the behavior can be fitted with $R_{NL} = \frac{1}{2}\gamma^2 \rho \frac{w}{\lambda_S} e^{-L/\lambda_S}$ where $\gamma$ is the spin Hall coefficient and $\lambda_S$ is the spin relaxation length [24]. With this, we found $\gamma$ to be ~ 0.2 and ~ 0.92 and $\lambda_S$ to be ~ 0.8 and ~ 0.34 μm for *FG1* and *FG2* respectively. The enhanced $\gamma$ and reduced $\lambda_S$ in *FG2* compared to less fluorinated *FG1* device is a direct indication of the larger spin-orbit strength in the former. Such small $\lambda_S$ values are in the expected range for low mobility graphene devices. However, unlike the case of hydrogen, large $\gamma$ value especially for FG2 is surprisingly large and doubtful. While we ruled out all the possible sources of leakage currents in our measurements, such large spin Hall angle might be due to the analysis. The equation we used above to estimate the SOC strength assumes that $l_e << w << \lambda_s$ where $l_e$ (~36 nm in FG2) is the electron mean free path, and $\lambda_s$ (~ 340 nm in FG2) is the electron spin relaxation length. Since it is very difficult to satisfy this condition experimentally, spin Hall angle values might not be truly extracted with the existing theory. Similar to our case, a recent study using the same theory for extracting the spin Hall angle found much larger values than the theoretically predicted values[27]. Further theoretical studies that satisfy the experimental conditions are required for the more accurate values.

In order to further quantity the devices, we now calculate the strength of SOC in fluorinated graphene devices. Following similar arguments used in hydrogenated graphene devices, we assume Elliott-Yafet type spin scattering mechanism as the dominant dephasing mechanism [8,9,23,28]. With this, we have $\Delta_{SOC} = E_F \sqrt{\tau_P/\tau_S}$, where $E_F$ is the Fermi energy, $\tau_P$ and $\tau_S$ are the momentum and spin relaxation times, respectively. We obtain a SOC of 5.1 and 9.1 meV for *FG1* and *FG2* samples at n = 1 x $10^{12}$ cm$^{-2}$, respectively. The observation of larger SOC in *FG2* compared to *FG1* is likely due to the presence of higher fluorination coverage. We note that the SOCs extracted for *FG1* and *FG2* device are nearly two and four times of what has been extracted for hydrogenated graphene devices at similar carrier concentrations, respectively [8]. In order to study the origin of this large SOC in fluorinated graphene compared to its hydrogenated counterparts, we estimate the out of plane distortion angle by assuming all SOC is caused by the *sp³* hybridization of carbon atoms. The distortion angles for *FG1* and *FG2* devices are extracted to be 9.9° and 19.2° for *FG1* and *FG2* samples, respectively by using $\emptyset = Arc\tan\left\{\left[1/4 - (9 - 8r_0^2)^{1/2}/12\right]^{1/2}\right\}$ [9]. The obtained 19.2° in our weakly fluorinated devices is surprising since the distortion angle for even a full *sp³* hybridization is 19.5° [29]. This implies that such large SOC cannot be explained only by the lattice deformation and hence there are additional contributions to the SOC in fluorinated graphene. In fact, recently it has been discussed independently by few studies that the fluorinated graphene should have larger SOC compared to the hydrogenated graphene due to the intrinsic SOC of the fluorine atoms [2,17]. Since the obtained SOC values in this study are comparable to the atomic spin orbit interaction in carbon and much higher than the hydrogenated graphene case[9], SOC in fluorinated graphene is not only from the graphene lattice but likely from the fluorine's own SOC[17]. The extracted spin parameters in these devices are summarized in Table1. We also extracted the spin parameters in a reference pristine-graphene based spin valve device for comparison.

In summary, we observe SHE in devices made from fluorinated graphene due to the enhancement of spin-orbit interaction in this defected system while, preserving high mobility of graphene. Large non-local signals are detected even without applying any external magnetic field. Observation of very strong spin orbit interactions in the order of 9 meV cannot be explained only with the out of plane distortion of the carbon bonds which is the main SOC source in hydrogenated graphene. Based on the recent theories, we believe that the spin orbit strength of fluorine adatom itself is the main source of the enhancement in our devices. The elimination of ferromagnetic contacts in these devices is a major advantage for the development of graphene-based spintronics applications.

B. Ö. would like to acknowledge support by the National Research Foundation, Prime Minister's Office and the SMF-NUS Research Horizons Award 2009-Phase II.

**Figure and Table Captions**

FIG. 1. (a) Optical image of a completed device with multiple Hall bar structures. (b) Raman intensity maps of the D, G and 2D bands for the device shown in (a). (c) Device schematics for the local and non-local measurement configurations

FIG. 2. (a) Raman spectrum study of pristine graphene, *FG1* and *FG2* devices. (b) Resistivity of pristine graphene, *FG1* and *FG2* devices as a function of carrier concentration at room temperature

FIG. 3. (a) Non-local resistance measurement of *FG2* device for $l/w = 2.5$ as a function of carrier concentration. Inset: Non-local measurement of pristine graphene device. $l/w$ is 2 in this device (b) Non-local resistance *FG2* device measured as a function of carrier concentration at different $l/w$. (c) Non-local resistance at $n=1\times10^{12} cm^{-2}$ as a function of length for *FG1* and *FG2*.

TABLE1. Extracted spin parameters of *FG1* and *FG2* devices at carrier concentration of $1 \times 10^{12}$ $cm^{-2}$. Spin parameters of a pristine-graphene measured with ferromagnetic electrodes are also included as reference.

# Figures

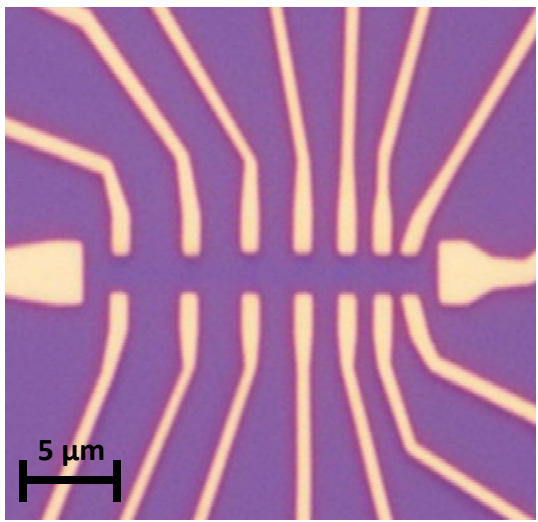 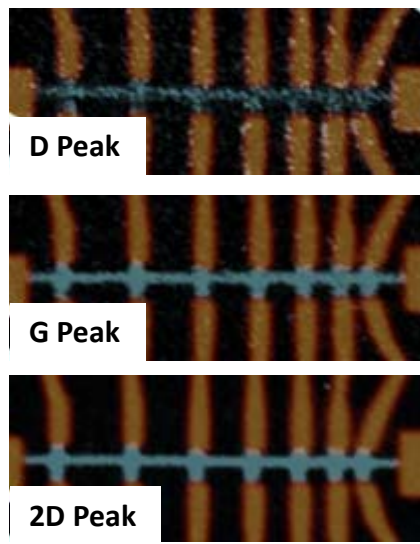 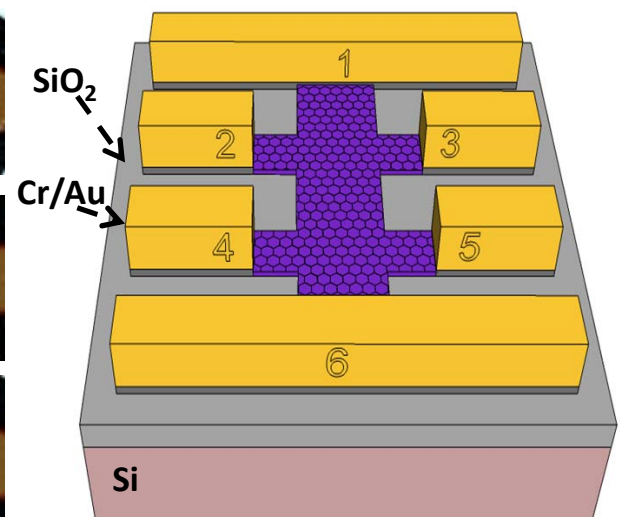

**Figure 1**

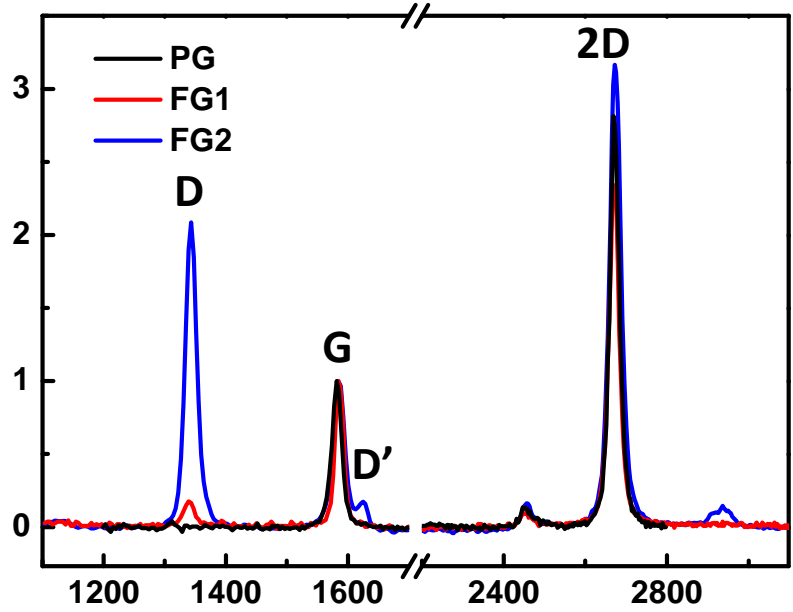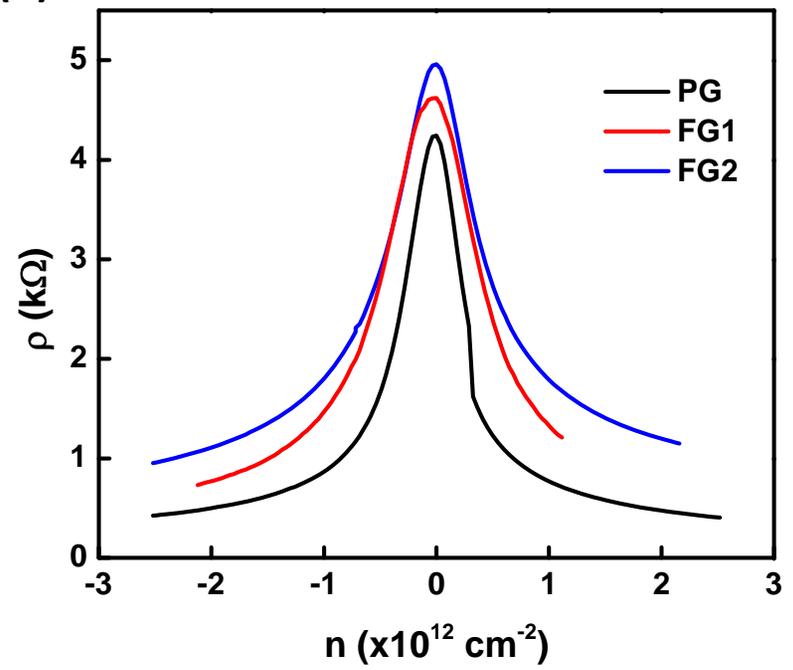

Figure 2

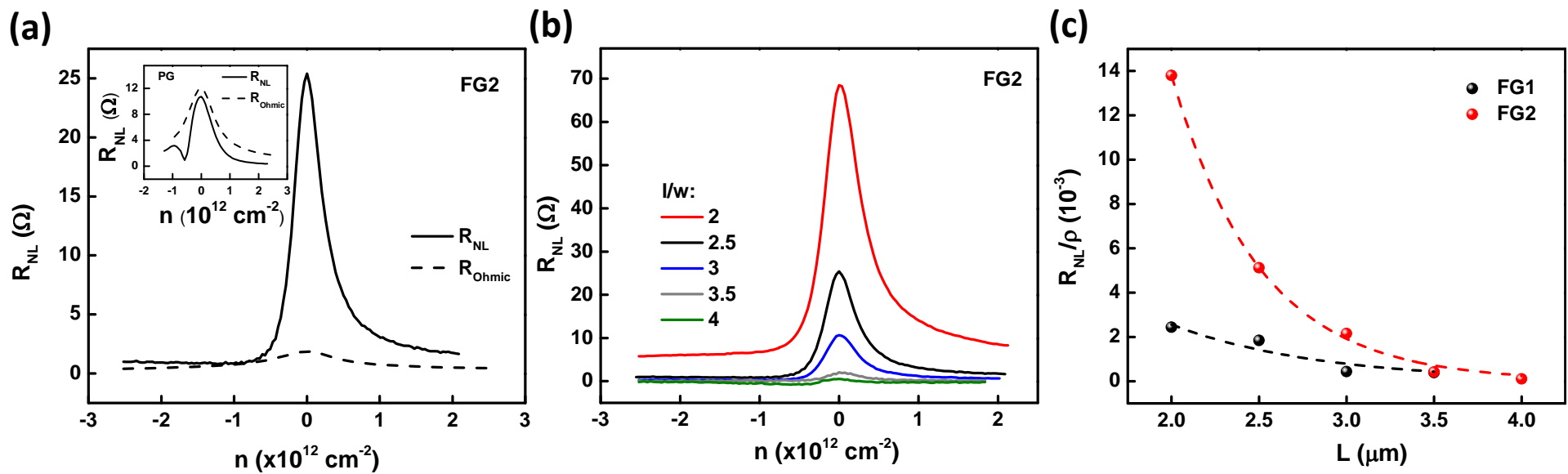

Figure 3

**Table 1**

| Sample | μ (cm²/Vs) | $\tau_S$ (ps) | $\lambda_S$ (μm) | $\Delta_{SOC}$ (meV) | θ (degree) |
|---|---|---|---|---|---|
| *FG1* | 4,600 | 25.8 | 0.8 | 5.1 | 9.9 |
| *FG2* | 2,700 | 5.9 | 0.34 | 9.1 | 19.2 |
| *Pristine-SV* | 2,900 | 135 | 1.5 | | |